\documentclass{article}

\usepackage{arxiv}

\usepackage[utf8]{inputenc} 
\usepackage[T1]{fontenc}    
\usepackage{hyperref}       
\usepackage{url}            
\usepackage{booktabs}       
\usepackage{amsfonts}       
\usepackage{nicefrac}       
\usepackage{microtype}      
\usepackage{lipsum}		
\usepackage{graphicx}
\usepackage{natbib}
\usepackage{doi}

\usepackage[cmex10]{amsmath}
\usepackage{booktabs}
\usepackage{multirow}
\usepackage{multicol}
\usepackage{hyperref}
\usepackage{lipsum}
\usepackage[table]{xcolor}

\title{Evaluation of Barlow Twins and VICReg self-supervised learning for sound patterns of bird and anuran species}

\author{ \href{https://orcid.org/0000-0003-2081-1233}{\includegraphics[scale=0.06]{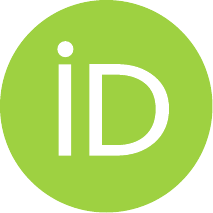}\hspace{1mm}Fábio Felix ~Dias}\\ 
    New York University\\
	\texttt{ffd2011@nyu.edu} \\
	\And
	\href{https://orcid.org/0000-0003-2059-9463}{\includegraphics[scale=0.06]{figs/orcid.pdf}\hspace{1mm}Moacir Antonelli ~Ponti} \\
	University of S{\~a}o Paulo\\
	\texttt{moacir@icmc.usp.br} \\
 	\And
	\href{https://orcid.org/0000-0002-4312-202X}{\includegraphics[scale=0.06]{figs/orcid.pdf}\hspace{1mm}Mílton Cezar ~Ribeiro} \\
	S{\~a}o Paulo State University\\
	\texttt{milton.c.ribeiro@unesp.br} \\
  	\And
	\href{https://orcid.org/0000-0002-4799-8774}{\includegraphics[scale=0.06]{figs/orcid.pdf}\hspace{1mm}Rosane ~Minghim} \\
	University College Cork\\
	\texttt{rosane.minghim@ucc.ie} \\
}

\date{}

\renewcommand{\shorttitle}{Evaluation of Barlow Twins and VICReg self-supervised learning for sound patterns}

\hypersetup{
pdftitle={Evaluation of Barlow Twins and VICReg self-supervised learning for sound patterns of bird and anuran species},
pdfsubject={cs.SD},
pdfauthor={Fábio Felix ~Dias, Moacir Antonelli ~Ponti, Mílton Cezar ~Ribeiro, Rosane ~Minghim},
pdfkeywords={sound event identification, animal species classification, classification improvement},
}

\begin{document}
\maketitle

\begin{abstract}
Taking advantage of the structure of large datasets to pre-train Deep Learning models is a promising strategy to decrease the need for supervised data. 
Self-supervised learning methods, such as contrastive and its variation are a promising way towards obtaining better representations in many Deep Learning applications. Soundscape ecology is one application in which annotations are expensive and scarce, therefore deserving investigation to approximate methods that do not require annotations to those that rely on supervision. Our study involves the use of the methods Barlow Twins and VICReg to pre-train different models with the same small dataset with sound patterns of bird and anuran species. In a downstream task to classify those animal species, the models obtained results close to supervised ones, pre-trained in large generic datasets, and fine-tuned with the same task.
\end{abstract}


\section{Introduction}
\label{sec:introduction}

Sound is an important attribute to understanding landscape dynamics~\cite{Pijanowski:2011a} and represents a relevant feature when investigating Machine Learning~\cite{Kong:2020} in the context of soundscape ecology.  
In that sense, recent research has focused on defining well-suited approaches to deal with natural sound tasks~\cite{Dufourq:2022,Stowell:2022}.
Those strategies are concentrated on Convolutional Neural Networks (CNNs) because they can recognize temporal-frequency patterns~\cite{Salamon:2017} and obtain proper results in many sound-related tasks~\cite{Kahl:2021,Lebien:2020}.

However, CNNs are data-hungry and the acquisition of large and accurately labeled datasets is difficult and time-consuming requiring alternative training strategies~\cite{ponti2021training}.
To handle this issue, Deep Learning specialists and researchers are attempting to take advantage of a large amount of unlabeled data available throughout the Internet with Self-supervised learning (SSL)~\cite{Sa:1994}.
For example,~\cite{Baevski:2020} developed a process capable of learning both contextualized speech representation and discretized speech units; and ~\cite{Saeed:2021} created a framework to learn general-purpose representations of speech sounds.

In this paper, we conducted experiments with contrastive methods, a popular SSL approach that minimizes the distance between similar patterns while maximizing the distance between unrelated patterns~\cite{Saeed:2021}.
Instead of using large unlabeled datasets, such as the aforementioned examples, we mainly evaluated the behavior of these techniques in cases in which the pre-training dataset has the same samples as the downstream task but with different views.
Even in such a restricted pre-training scenario, SSL techniques pre-trained models that reached results near to models pre-trained on well-known supervised tasks with large datasets.
The best models and codes will be available on GitHub\footnote{\url{https://github.com/fabiofelix/Sound-Self-Supervised}}.

The subsequent sections are constructed as follows.
\autoref{sec:concepts} concisely reviews the SSL approaches tested in this research.
\autoref{sec:method} presents the steps followed and the materials used in our experiments.
\autoref{sec:results} reports the experimental results obtained with the experiments.
\autoref{sec:discussion} discusses the experimental results.
Finally, \autoref{sec:conclusion} provides the conclusions and directions for future work.

\section{Technical Background}
\label{sec:concepts}

With SSL strategies, it is possible to use auxiliary tasks, whose labels are created from the dataset structure, and the learned weights are transferred to specific tasks, obtaining suitable results in many real applications~\cite{Baevski:2020,Chi:2021,Owens:2018}.

One recurrent SSL technique pre-trains a model with branches that learn similar invariant embeddings of different data views~\cite{Bardes:2022}.
As an example, Barlow Twins~\cite{Zbontar:2021} has a structure with two consecutive blocks, an \emph{encoder} and a \emph{projector}.
The former can be any CNN without the classification layer and the latter has three dense layers with $N$ units, \emph{batch normalization} before ReLU activation (two first layers), and the last layer only with linear activation.
Barlow Twins authors claim that the \emph{projector} removes space redundancy of the \emph{encoder}.
This two-block structure configures each branch of a Siamese network~\cite{Bromley:1994,Chopra:2005} with shared weights between the branches. Such a method shows good representation learning ability while maintaining robustness to attack scenarios~\cite{cavallari2022training}.
The architecture receives two different views of the input and forces their embeddings to be close, minimizing component redundancy of the learned features by employing the following contrastive loss function:

\begin{equation}
  L_{BT} = \underbrace{\sum_i (1 - C_{ii})^2}_{\text{invariance term}} + \underbrace{\lambda \sum_i\sum_{i\neq j}C_{ij}^2}_{\text{redundancy reduction term}}\text{,}
\end{equation}

\noindent being $\lambda$ a hyperparameter that controls the term importance and $C$ the cross-correlation matrix of the spaces of the network branches.
The invariance term forces the \emph{encoder} spaces to be close and the second term reduces the spaces' redundancy.
Besides, it can avoid space collapse to constant values or irrelevant data information. 
After the training, the \emph{encoder} is extracted and fine-tuned in a downstream task.  

As a variation of this idea, \cite{Bardes:2022} also defined the Variance-Invariance-Covariance Regularization (VICReg) to avoid space collapse.
This regularization technique can stabilize the training process, leading to results close to the state-of-the-art in many downstream tasks.
We can use it with the Barlow Twins configuration but with the following contrastive loss function: 

\begin{equation*}
\begin{split}
  \ell(Z_a, Z_b) & = \underbrace{\lambda \frac{1}{n} \sum mse(Z_a, Z_b)}_{\text{invariance}} + \underbrace{\mu\left[ v(Z_a) + v(Z_b) \right]}_{\text{variance}} + \underbrace{\nu\left[ c(Z_a) + c(Z_b) \right]}_{\text{covariance}}\text{,}
\end{split}
\end{equation*}

\noindent being $Z_i$ a data \emph{batch} with $n$ vectors learned by the \emph{expander} (\emph{projector}) of each network branch, and $\lambda$, $\mu$, and $\nu$ hyperparameters.
The invariance term is the \emph{mean squared error} (mse) between the spaces, and the variance and covariance are defined as:

\begin{multicols}{3}
    \begin{equation*}
      v(Z) = \frac{1}{n} \sum relu(\gamma - \sqrt{var(Z) + \epsilon})\text{,}
    \end{equation*}

  \begin{equation*}
    c(Z) = \frac{1}{N} \sum_i\sum_{i\neq j}C(Z)_{ij}^2\text{,}
  \end{equation*}

  \begin{equation*}
    C(Z) = \frac{\left\lVert Z - mean(Z)\right\rVert_2}{N - 1}\text{,}
  \end{equation*}    
\end{multicols}

\noindent where $N$ is the space size of the \emph{expander}, $\gamma = 1$, $\epsilon = 10^{-4}$, and $mean$ and $var$ are calculated for the \emph{batch}.
The invariance term has the same purpose as in the Barlow Twins, the variance term does not allow the vectors to point to the same place, and the covariance term reduces the redundancy of the learned vectors.

\section{Method}
\label{sec:method}

This section presents the steps to evaluate SSL strategies to pre-train CNNs used to identify animal species.
In the first step, the training subset of~\autoref{tab:dataset} was balanced by data augmentation strategies for sound signals.
We used these same techniques to create different views to feed the SSL tasks.
During the second step, we built a baseline by fine-tuning models to identify sound patterns with weights randomly initialized or pre-trained on generic image classification tasks.
The third step consists of pre-training models with SSL tasks and fine-tuning them to also identify sound patterns.

\subsection{Dataset}
\label{sec:dataset}

Our dataset in~\autoref{tab:dataset} contains recordings collected on natural landscapes, provided by the Spatial Ecology and Conservation Lab (LEEC) \footnote{\url{https://github.com/LEEClab}}, and already explored by~\cite{Scarpelli:2021,Hilasaca:2021b,Hilasaca:2021a,Dias:2021b}.
These data are part of the Long Term Ecological Research conducted at the Cantareira-Mantiqueira ecological corridor (LTER CCM), which is localized in the northeast portion of Sao Paulo state, Brazil. Moreover, following~\cite{Kahl:2021}, the table has samples from Google AudioSet~\cite{Gemmeke:2017}, forming a dataset with 15 classes.
To download AudioSet recordings, we employed youtube-dl (v2021.4.26) library.

The dataset was split using a stratified method of the classes in training (90\%) and test (10\%), totaling 5000 clips of 3 seconds (250 min. in total), as showed in~\autoref{tab:dataset}.
We applied $k$-fold cross-validation with $k = 5$ in the training subset and for each iteration, one partition was used as a validation subset.
In that sense, fine-tuning tasks used training, validation, and test subsets, meanwhile, SSL auxiliary tasks used only the training and validation subsets.


\begin{table}[!h]
  \caption{Quantities of 3-second audio clips grouped by species and named with a short label}
  \label{tab:dataset}    
  \centering
   \begin{tabular}{lllccc}  
  \toprule
                          & \textbf{specie}     &  \textbf{label}     & \textbf{\#train} & \textbf{\#test} & \textbf{Total} \\
  \cmidrule(lr){1-3}  
  \cmidrule(lr){4-6}
  \multirow{6}{*}{bird}   & \emph{Basileuterus culicivorus}  & basi\_culi & 483        & 54  &  537  \\
                          & \emph{Cyclarhis gujanensis}      & cycl\_guja & 390        & 43  &  433  \\
                          & \emph{Myiothlypis leucoblephara} & myio\_leuc & 411        & 46  &  457  \\
                          & \emph{Pitangus sulphuratus}      & pita\_sulp & 352        & 39  &  391  \\
                          & \emph{Vireo chivi}               & vire\_chiv & 724        & 81  &  805  \\
                          & \emph{Zonotrichia capensis}      & zono\_cape & 574        & 64  &  638  \\
  \cmidrule(lr){4-6}
          & &  & 2934      & 327 & 3261  \\
  \cmidrule(lr){1-3}  
  \cmidrule(lr){4-6}
  \multirow{6}{*}{anuran} & \emph{Adenomera marmorata}        & aden\_marm & 116        & 13  &  129  \\
                          & \emph{Aplastodiscus leucopigyus}  & apla\_leuc & 186        & 21  &  207  \\
                          & \emph{Boana albopunctata}         & boan\_albo & 283        & 32  &  315  \\
                          & \emph{Dendropsophus minutus}      & dend\_minu & 229        & 26  &  255  \\ 
                          & \emph{Ischnocnema guenteri}       & isch\_guen & 136        & 15  &  151  \\
                          & \emph{Physalaemus cuvieri}        & phys\_cuvi & 290        & 32  &  322  \\
  \cmidrule(lr){4-6}
          &  &  & 1240      & 139 &  1379  \\  
  \cmidrule(lr){1-3}  
  \cmidrule(lr){4-6}
  \multirow{3}{*}{other}  & & animal    & 108         & 12  &  120  \\
                          & & human     & 109         & 11  &  120  \\
                          & & natural   & 109         & 11  &  120  \\
  \cmidrule(lr){4-6}
        &  &  & 326         & 34  &  360   \\    
  \cmidrule(lr){1-3}  
  \cmidrule(lr){4-6}
  \textbf{Total} & & & 4500   & 500 & 5000 \\
  \bottomrule
  \end{tabular}
\end{table}

\subsection{Balancing data classes}
\label{sec:augmentation}

The dataset was augmented both to reduce possible problems, such as model poor generalization and improper predictions for samples of minority classes~\cite{Johnson:2019,Wang:2017b}, and to create different data views for SSL tasks.

To fine-tune model weights, augmentation generated $m=\lceil(\max_c -\#class)/\#class\rceil$ modified copies of all 3-second training recordings that were added to the original training subset to generate spectrograms.
When cross-validation split training files, we first divide the originals into 3600 files ($k - 1$ partitions) for training and 900 samples for validation.
After that, we take augmentations until each class in the training subset reaches $\max_c=580$ audio clips, generating a set with 8700 clips.
The choice of 580 is to approach the majority class (\textit{Vireo chivi}) in the training subset.
Hence, in each cross-validation iteration, there are 8700 clips for training, 900 clips for validation, and 500 clips for testing.

We considered \emph{time stretching}, \emph{pitch shifting}, and \emph{noise addition} as proposed by~\cite{Salamon:2017}, following the same implementations used in~\cite{Dias:2021b} and parameters of~\autoref{tab:auto_aug_param}.

\begin{table}[!ht]
  \caption{Values range for the augmentation parameters.}   
  \label{tab:auto_aug_param}    
  \centering    
  \begin{tabular}{|l|l|}
    \hline    
                              & \textbf{values} \\
    \hline                            
    \textit{stretch} (factor) & from 0.7 to 1.3, incremented by 0.1 (discards 0.0)  \\
    \textit{pitch}  (steps)   &  $[-12, -6, -3, 3, 6, 12]$            \\
    \textit{noise} (dB)       &  from 2 to 12, incremented by 2 \\ 
    \hline        
  \end{tabular}
\end{table}

For SSL tasks, augmentation creates two alternative views for each training and validation subset sample.
The process randomly chooses two functions and their respective parameters described in~\autoref{tab:auto_aug_param}.
After cross-validation splits data in training and validation, equal to the past description, the augmentation generates alternative views and each iteration has 7200 samples for training and 1800 for validation.

\subsection{Audio spectrograms}

For all audio clips, gray-scale mel-spectrogram images (256$\times$256) were created with librosa (v0.8.1) routines, using a Hanning window with a length of 2048 and an overlap of 75\%.
Length and overlap contribute to building a representation with suitable frequency and time resolutions to represent a large pattern variation.

\subsection{Network architectures}
\label{sec:archtecture}

We employed four network architectures: one proposed by~\cite{Salamon:2017}, named here as SimpleCNN, MobileNet-V3 (Large)~\cite{Howard:2019}, ResNet-50~\cite{He:2016}, and Inception-V3~\cite{Szegedy:2016}, all of them pre-trained on ImageNet~\cite{Deng:2009} when necessary.
SimpleCNN achieved suitable results to classify bird species, MobileNet is an architecture built to consume less computational resources, ResNet is a common option to classify animal species~\cite{Harvey:2018,Lebien:2020,Thomas:2019}, and Inception has filter variations that could improve the learning process of patterns with different sizes.

To train SimpleCNN and MobileNet-V3, we used SGD optimizer with learning rate $lr = 10^{-2}$ and $momentum = 0.9$ (only MobileNet-V3).
ResNet-50 used Adam with $lr = 10^{-4}$ and Inception-V3 used RMSProp with $lr = 10^{-3}$.
All training processes executed 100 epochs with $batch = 80$, except ResNet-50, whose $batch = 30$.

SSL tasks and fine-tuning used the same configurations, changing the batch size because of memory constraints: 30 samples for ResNet-50 and 50 samples for the others.
All inputs were normalized with \emph{max-norm}, dividing pixels values by 255.

We created a SimpleCNN version pre-trained on a small version of ImageNet.
We used $\approx 28\%$ (358.727 samples) from the training subset and maintained the samples of validation (50K samples) and test (100K samples).
This pre-training considered the same optimization setup described in the preceding paragraphs with $batch = 256$.

Models were implemented with Python (v3.8.10) associated with TensorFlow/Keras (v2.7.0) library.

\subsection{Self-supervised tasks}

We considered the Barlow Twins and VICReg described in~\autoref{sec:concepts}.
Their inputs are mel-spectrograms of the views generated by the~\autoref{sec:augmentation} and the models are initialized with random weights.
The configuration of the respective loss functions follow the hyperparameters defined in the original papers: Barlow Twins $\lambda = 0.005$; and VICReg $\lambda = \mu = 25$, $\nu = \gamma = 1$, and $\epsilon = 10^{-4}$.

The \emph{encoder} of both tasks are configured with one of the cited architectures, discarding the SimpleCNN dense layers, and replacing the top layers of the other networks with \emph{global average pooling}.
Furthermore, the three dense layers of the \emph{projector} have $N = 512$.

After pre-training, we extracted the \emph{encoder} of the task that achieved the best loss function value in the validation subset.
Besides, we added the top layers of each model (randomly initialized) and fine-tuned the learned weights.
 
\subsection{Evaluation}

We evaluated the results with the balanced accuracy score of scikit-learn (v1.0.1)~\cite{scikit-learn:2011}.
We also employed a two-tailed Student's \emph{t}-test to compare results using paired tests with a significance level equal to 0.05. 

In all tests, we used Python seed values (1030), following the Keras FAQ\footnote{\url{https://keras.io/getting_started/faq/}}.
As aforesaid, we used cross-validation with $k = 5$ in both cases pre-train and fine-tune.
For SSL pre-training, we evaluated the loss function in the validation subset, and for fine-tuning we considered the mean and standard deviation of the balanced accuracy of the test subset.

To perform comparisons, we executed three cases: in the first, all weights were randomly initialized; in the second, with weights learned on ImageNet; and finally, with weights learned with SSL tasks.
We fine-tuned models to classify 15 classes, but we assessed them with twelve classes related to the specific animal species of interest (see~\autoref{tab:dataset}), to facilitate comparisons with previous works.

Besides, we evaluated the impact of changing the number of samples: \emph{i)} doubling the training subset with more unlabeled samples from LEEC data to pre-train with SSL tasks, \emph{ii)} and diminishing the training subset to fine-tune the models to 50\% and 20\% of the training subset in~\autoref{tab:dataset}.
We also assessed variations of the \emph{projector} space size $N\in[512, 1024, 2048, 4096, 8192]$ to compare with the results of~\cite{Zbontar:2021,Bardes:2022}.
We tested only with the best combination of model and SSL task (ResNet50 and VICReg) and used the same configurations of~\autoref{sec:archtecture} to pre-train the model, with $batch = 90$ (smallest $N$), $batch = 80$ (largest $N$).

Associated with balanced accuracy, we employed silhouette coefficient~\cite{Tan:2005} to numerically evaluate learned features and t-SNE~\cite{Maaten:2008} that visually helps the evaluation of classes' neighborhoods and their segregation~\cite{Nonato:2018}.
To perform this, we extracted features from the penultimate layer because they are the inputs for the classification layer.
The routines to generate silhouettes and projections are also available on scikit-learn.

Finally, the majority of training and tests were performed with an NVidia Titan XP video card, with driver v470.86, Cuda v11.2.152, and cuDNN v8.1.0.
The SimpleCNN pre-train on ImageNet and the variation of \emph{projector} space were executed with an NVidia RTX A5000 video card, with driver v470.74, Cuda v11.4, and cuDNN v8.1.0.

\section{Results}
\label{sec:results}

This section reports the results of our experiments with SSL tasks.
Such results were compared with those of the same models initialized with random weights and weights learned on ImageNet.
The evaluation uses balanced accuracy $\in [0, 1]$, silhouette coefficient $\in [-1, 1]$, and t-SNE projections.
We executed two SSL tasks: Barlow Twins and VICReg; four CNNs, named: SimpleCNN, MobileNetV3, ResNet50, and InceptionV3, all fed with mel-spectrogram images.
Finally, tests were executed with $5$-fold cross-validation.

\begin{table}[!h]
  \caption{ 
    Results of models randomly initialized, pre-trained on ImagetNet, and SSL tasks.
    Both silhouette coefficient and balanced accuracy were calculated for the test subset.
    Silhouette is related to the best cross-validation partition.
    Highlighted values are the best results.
  }
  \label{tab:autossupervisao_resultado}    
  \centering  
  \begin{tabular}{lllllllll}
  \toprule    
  & \multicolumn{2}{c}{ \textbf{Random} } & \multicolumn{2}{c}{ \textbf{ImageNet} }  \\
  \cmidrule(lr){2-3}
  \cmidrule(lr){4-5}
            & \textbf{silhouette} & \textbf{balan.acc.} & \textbf{silhouette} & \textbf{balan.acc.} \\
  \cmidrule(lr){2-3}
  \cmidrule(lr){4-5}
  SimpleCNN    &  0.0065  & 0.60{\tiny$\pm$ 0.02} &  0.0340  & 0.61{\tiny$\pm$ 0.02}  \\
  MobileNetV3 &  0.0863  & 0.62{\tiny$\pm$ 0.03} &  0.1500  & 0.70{\tiny$\pm$ 0.01}  \\
  ResNet50    &  0.0769  & 0.65{\tiny$\pm$ 0.03} &  \cellcolor[HTML]{CFFFFF}\textbf{0.2240}  & \cellcolor[HTML]{CFFFFF}\textbf{0.77{\tiny$\pm$ 0.01}} \\
  InceptionV3 &  0.1150  & 0.67{\tiny$\pm$ 0.03} &  0.1303  & 0.77{\tiny$\pm$ 0.02}  \\
  \midrule
  & \multicolumn{2}{c}{ \textbf{Barlow Twins} }  & \multicolumn{2}{c}{ \textbf{VICReg} }  \\
  \cmidrule(lr){2-3}
  \cmidrule(lr){4-5}
  SimpleCNN    & -0.0031  & 0.55{\tiny$\pm$ 0.02} & 0.0142   & 0.61{\tiny$\pm$ 0.01} \\
  MobileNetV3 & -0.1213  & 0.23{\tiny$\pm$ 0.07} & 0.0085   & 0.54{\tiny$\pm$ 0.03} \\
  ResNet50    & 0.1041   & 0.69{\tiny$\pm$ 0.03} & 0.1374   & 0.72{\tiny$\pm$ 0.02} \\
  InceptionV3 & 0.0451   & 0.68{\tiny$\pm$ 0.02} & 0.0109   & 0.68{\tiny$\pm$ 0.02} \\
  \bottomrule
  \end{tabular}
\end{table}

\begin{figure*}[!h]
  \centering
   \includegraphics[scale=0.62]{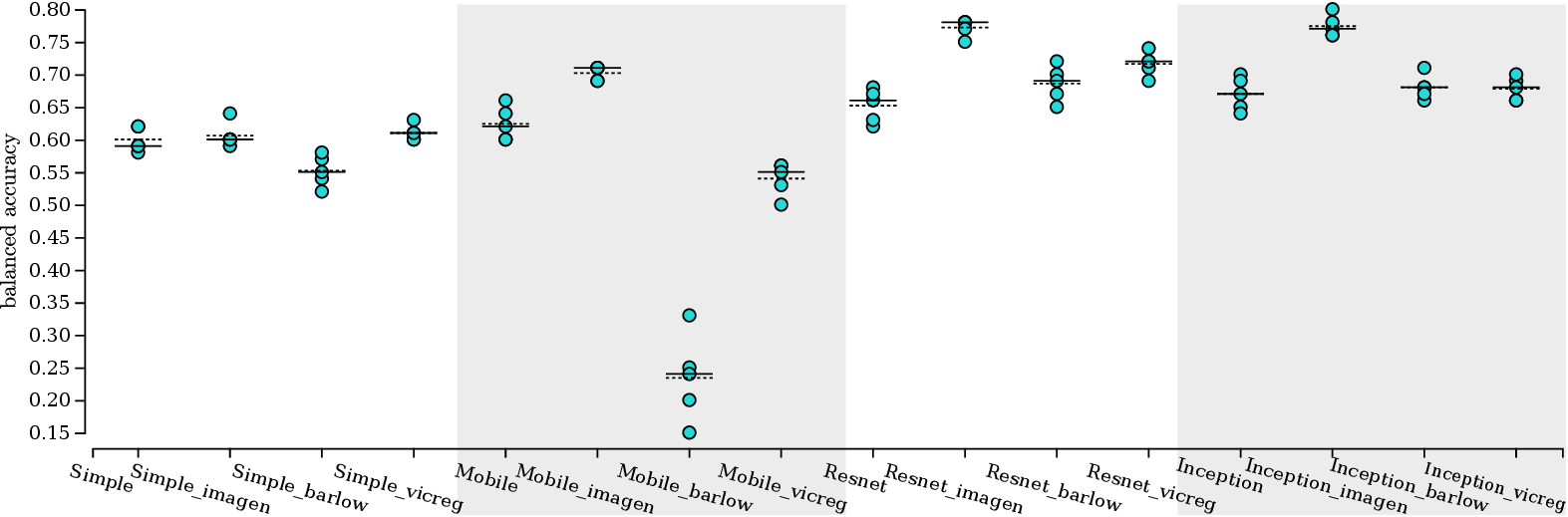} 
\caption{
    Individual value plots of balanced accuracy of CNNs applied to test subset.
    Each plot area (white and shade gray) shows the best results of a model with different weight initialization (from left to right: random, pre-trained on ImageNet, and pre-trained with SSL).  
    Solid and dotted lines in each group communicate median and mean values, respectively. 
    An overlap between points generates model results with a different number of points.}
  \label{fig:self_boxplot1}   
\end{figure*}

\autoref{tab:autossupervisao_resultado} and~\autoref{fig:self_boxplot1} summarize the results.
In all test cases, ResNet50 and InceptionV3 obtain similar results (do not reject the null hypothesis with p-value $> 0.05$) and are always superior to the other models, with more than 3 percent points of difference. 
ResNet50 also reached the highest silhouette coefficient in all pre-trained cases.

Random initialization and transfer from ImageNet are respectively lower and higher bounds of the results and the two SSL strategies reduced silhouette ($< 0.14$) and balanced accuracy ($\le 0.72$) results when compared with the higher bounds (silhouette $< 0.23$ and accuracy $\le 0.77$).
In that sense, MobileNet suffered the most with SSL pre-training, reducing the results to values below the random initialization.
The comparisons between Barlow Twins and VICReg show that the latter has results at least 3 percent points greater than the former SSL, except for InceptionV3.
These and other differences and similarities between the results are more visible when analyzing~\autoref{fig:self_boxplot1}. 

\subsection{Learned feature space}

\begin{figure*}[!h]
  \centering
   \includegraphics[scale=1.0]{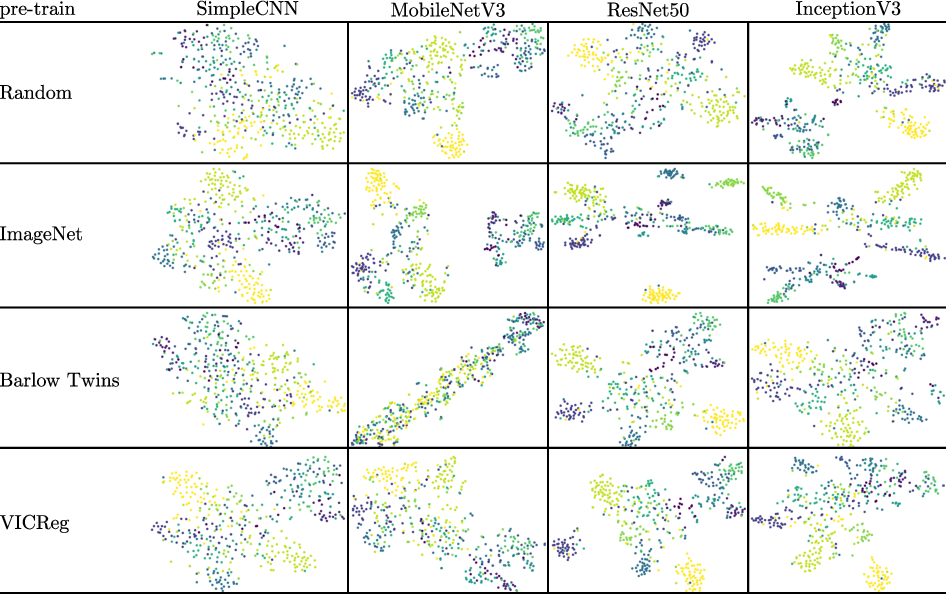} 
\caption{t-SNE projection of the test subset represented by model's learned features. 
    Each figure column shows the best results of a model with different weight initialization.  
    Models were chosen from the best cross-validation partition of each test case.
    Point colors represent species of interest.}
  \label{fig:self_projecao1}
\end{figure*}

\autoref{fig:self_projecao1} has t-SNE projections to inspect the learned feature spaces visually.
These images ratify the results described previously.
Overall, the model's depth impacts class segregation, being the features learned by SimpleCNN the ones that present more visual clutter, and ResNet50 and Inception the models that learned features with well-defined visual distinction.
Furthermore, the models pre-trained with ImageNet achieve the best visual segregation.
These behaviors reflect silhouette coefficients reported on~\autoref{tab:autossupervisao_resultado}. 
One can also verify that the innermost regions of the projections have more visual clutter than the outermost ones. 

We also colored data points with two colors (bird and anuran), and it was possible to verify, independently of the model or weight initialization, a clear visual segregation between the two groups, even with some overlap level.

\subsection{Variation of training samples}

As we could expect when reducing the training subset to fine-tune the models, 
the balanced accuracy diminished significantly (reject the null hypothesis with p-value $\le 0.05$).
For example, ResNet50 pre-trained with VICReg: 
$0.72\pm 0.02$ (refined with 100\% of training subset), $0.60\pm 0.02 $ (refined with  50\%), and $0.43\pm 0.02$ (refined with  20\%).

Doubling the number of samples to execute SSL increased the processing time from one day to three days.
This increment of samples impacts more Barlow Twins, increasing results in most cases.
For example, it increased ResNet50 results from $0.69\pm 0.03$  (see~\autoref{tab:autossupervisao_resultado}) to $0.71\pm 0.01$, and MobileNetV3 from $0.23\pm 0.07$ to $0.32\pm 0.08$.

In both tests, following the~\autoref{tab:autossupervisao_resultado}, VICReg results are greater than or equal to Barlow Twins results.

\subsection{Variation of self-supervision space}

In this test, we considered only ResNet50 pre-trained with VICReg because it achieves the best SSL results on~\autoref{tab:autossupervisao_resultado}.
The $N$ variations did not change significantly the results (not reject the null hypothesis with p-value $> 0.05$).
For example, with the smallest \emph{projector} space ($N = 512$) the fine-tuned ResNet50 obtains balanced accuracy of $0.70\pm 0.02$, while with the largest space ($N = 8192$, best results on original papers) obtains $0.69\pm 0.02$.

\section{Discussion}
\label{sec:discussion}

In general, random initialization of weights and weights transferred from ImagetNet tasks are lower bound (balanced accuracy mean $\in [0.60, 0.67]$) and upper bound (balanced accuracy mean $\in [0.61, 0.77]$) of the results.
Hence, depending on the model used as \emph{encoder}, models pre-trained with Barlow Twins achieved results close to random initialization, with maximum balanced accuracy equal to 0.69, while with VICReg, models achieved results up to 0.72.
Such behavior is similar to the results reported by~\cite{Bardes:2022}, which reports the superiority of VICReg results, highlighting its capability of producing more robust features.
Furthermore, even with a small train (4500 samples from 15 classes), VICReg obtained models that can yield 5 percent points away from the ImageNet pre-trained models (+1M samples from 1K classes).

In our tests, independently of the weight initialization approach, the refined models have difficulties yielding balanced accuracy greater than 0.77 and silhouette coefficient greater than 0.23.
This issue is related to the level of class overlap viewed in parts of~\autoref{fig:self_projecao1}.

As could be expected, the fewer data samples for fine-tuning, the lesser the balanced accuracy of the models.
However, models pre-trained with VICReg achieved the highest results than the ones pre-trained with Barlow Twins.
Overall, adding more samples for pre-training did not significantly increase the results, except for MobileV3, which increased results from $0.23\pm 0.07$ to $0.32\pm 0.08$.
Even so, it demands more tests to evaluate the impact of the size of the pre-train dataset on the final results.
Besides, increasing pre-training samples also increased 3x the processing time. 

Finally, variations of the \emph{projector} space size did not impact the results and 
the balanced accuracy became stable at around 0.69.  
This behavior is different from the results of~\cite{Zbontar:2021,Bardes:2022}, which reported an improvement in the results when increasing the space size.
Therefore we have to conduct more experiments to achieve a better understanding of it.   

\section{Conclusion}
\label{sec:conclusion}

This paper reported a series of tests with SSL to evaluate their impact on CNNs fine-tuned to identify natural sound patterns.
We have tested two SSL tasks and four architectures and compared the final results with the same models pre-trained on the ImageNet and initialized with random weights.
In our tests, the VICReg strategy pre-trained models that converge to better results than Barlow Twins, following~\cite{Bardes:2022}.
Furthermore, even with a small train dataset (4500 samples from 15 classes), it is possible to pre-train with VICReg and fine-tune models that achieve balanced accuracy up to 0.72, while the same models pre-trained with ImageNet (+1M samples from 1K classes) reach 0.77.

Changes on the SSL hyperparameter (space size) did not corroborate with their original papers and increments of the dataset for pre-training did not generate significant improvements but increased the processing time 3x.
Future work is necessary to understand variations of the dataset size, technique hyperparameter, and the applicability of other approaches, such as~\cite{Cramer:2019,Guzhov:2022}.

\section*{Acknowledgment}

This study was financed in part by the Coordenação de Aperfeiçoamento de Pessoal de Nível Superior - Brasil (CAPES) - Finance Code 001, FAPESP (grant 2019/07316-0 and 2021/08322-3), and CNPq (National Council of Technological and Scientific Development) grant 304266/2020-5. MCR thanks to the Sao Paulo Research Foundation - FAPESP (processes 2013/50421-2; 2020/01779-5; 2021/08322-3; 2021/08534-0; 2021/10195-0; 2021/10639-5;  2022/10760-1) and National Council for Scientific and Technological Development - CNPq (processes 442147/2020-1; 440145/2022-8; 402765/2021-4; 313016/2021-6; 440145/2022-8), and Sao Paulo State University - UNESP for their financial support. This study had the collaboration of infrastructure and support from the Pierre Kaufmann Observatory Radio, located in the city of Atibaia, Sao Paulo. The Pierre Kaufmann Radio Observatory is an institution maintained and operated by the Mackenzie Presbyterian University through its Mackenzie Radio Astronomy and Astrophysics Center (CRAAM) in collaboration with the National Institute for Space Research (INPE). We kindly thank Guilherme Alaia for all the support that he gives to the LTER CCM team. This study is also part of the Center for Research on Biodiversity Dynamics and Climate Change, which is financed by the Sao Paulo Research Foundation - FAPESP.

\bibliographystyle{unsrtnat}
\bibliography{references,references_doc,references_mestrado}  






\end{document}